\documentclass[aps,prb,twocolumn,superscriptaddress]{revtex4-1}

\usepackage[dvips]{graphicx}
\usepackage[dvips]{color}
\usepackage{amsmath}
\usepackage{amssymb}
\usepackage{multirow}

\usepackage{bm}

\renewcommand{\Vec}[1]{\mbox{\boldmath$#1$}}

\def\t#1{\textrm{#1}}
\def\n{\nonumber \\ }
\def\tensor{\otimes}
\def\Z{\mathbb{Z}}

\begin{document}

\title{Weyl and Dirac semimetals
with $\mathbb{Z}_2$ topological charge}

\author{Takahiro Morimoto}
\affiliation{Condensed Matter Theory Laboratory, RIKEN, Wako, Saitama, 351-0198, Japan}
\author{Akira Furusaki}
\affiliation{Condensed Matter Theory Laboratory, RIKEN, Wako, Saitama, 351-0198, Japan}
\affiliation{RIKEN Center for Emergent Matter Science (CEMS), Wako, Saitama, 351-0198, Japan}
\date{\today}

\begin{abstract}
We study the stability of gap-closing (Weyl or Dirac) points in the
three-dimensional Brillouin zone of semimetals using
Clifford algebras and their representation theory.
We show that a pair of Weyl points with $\Z_2$ topological charge
are stable in a semimetal with time-reversal and reflection symmetries
when the square of the product of the two symmetry transformations
equals minus identity.
We present toy models of $\Z_2$ Weyl semimetals which have
surface modes forming helical Fermi arcs.
We also show that Dirac points with $\Z_2$ topological charge
are stable in a semimetal with time-reversal, inversion, and
SU(2) spin rotation symmetries when the square of the product of
time-reversal and inversion equals plus identity.
Furthermore, we briefly discuss the topological stability of
point nodes in superconductors using Clifford algebras.

\end{abstract}


\maketitle

\section{Introduction}
Weyl semimetals\cite{murakami-semimetal07,Wan-semimetal11,burkov-balents11,burkov-hook-balents11,Yang-Lu-Ran11,halasz-balents12,zyuzin12,fang12,bulmash14,volovik2003universe,jianhua-jiang12,teemu13,okugawa-murakami14}
are three-dimensional (3D) analogs of graphene
and have gapless low-energy excitations of Weyl fermions.
The low-energy effective Hamiltonian for Weyl fermions
has the form
\begin{align}
H_0=k_x \sigma_x + k_y \sigma_y + k_z \sigma_z,
\end{align}
where the Fermi velocity is set to unity and
the wave number $\bm{k}$ is measured from a Weyl point.
Since all three Pauli matrices $\sigma_\alpha$ ($\alpha=x,y,z$) are exhausted
by the kinetic terms in the low-energy Hamiltonian,
the Weyl fermions are massless and stable against perturbations.
The stability of Weyl points has a topological origin.
For any fixed value of $k_z\,(\ne0)$ at which
the energy band structure is gapped,
a Chern number $\nu(k_z)$ can be defined
on the two-dimensional (2D) $k_x$-$k_y$ plane.
As $k_z$ is varied, 
$\nu(k_z)$ can change only when the 2D $k_x$-$k_y$ plane
crosses a Weyl point.
We can thus assign to each Weyl point an integer ($\mathbb{Z}$) topological
charge which is the change in $\nu(k_z)$ at
the topological phase transition.
The well-defined topological charge makes Weyl points stable.
A nontrivial value of the Chern number $\nu(k_z)$ also guarantees
that there exist chiral surface states
which form a Fermi arc
connecting projections of
two Weyl points with opposite charges onto the
surface Brillouin zone.
However, the topological stability of Weyl points is lost when
both time-reversal and inversion symmetries are present in the material,
because the combination of the two symmetries constrains two Weyl points
with opposite Chern numbers to merge, thereby making the total
topological charge vanish.\cite{murakami-semimetal07,burkov-balents11,halasz-balents12}

A natural question we may ask is whether
there are $\Z_2$ analogs of Weyl semimetals,
in a similar way to the way
we have 2D $\Z_2$ topological insulators\cite{kane-mele05,kane2005z_}
as opposed to integer quantum Hall systems characterized by a Chern number.\cite{TKNN}
In this paper, we propose two kinds of $\Z_2$ semimetals
which are topologically stable in the presence of time-reversal symmetry and
additional spatial symmetry.
First, we show that semimetals with a pair of Weyl points
characterized by $\Z_2$ topological charge are stable in the
presence of both time-reversal symmetry and (a kind of)
reflection symmetry which we define later.
In this semimetal, which we dub \textit{$\mathbb{Z}_2$ Weyl semimetal},
we can define a $\Z_2$ topological number for any 2D
cut of the Brillouin zone which is
parallel to the reflection plane 
and away from Weyl points.
Helical edge modes exist on the 2D cut with a nontrivial $\Z_2$
topological number, and a 2D surface perpendicular to the
reflection plane has helical Fermi arcs in the surface Brillouin zone.
Second, we show that Dirac semimetals having stable Dirac points
with $\Z_2$ topological charge are possible
in materials with SU(2) spin rotation, time-reversal,
and inversion symmetries.
We shall call this class of semimetals \textit{$\Z_2$ Dirac semimetals}.
In Table~\ref{table: topological charges},
we summarize topological charges of gap-closing points in semimetals
under given symmetries.
The type of topological charges depends on the sign of squares of
symmetry operators, or equivalently
commutation/anticommutation relations between symmetry operators.
For example, $\Z_2$ Weyl semimetals with time-reversal and
``reflection'' symmetries are stabilized under reflection symmetry
operator $R_z$ that squares to $+1$ and
commutes with time-reversal symmetry operator $T$ ($T^2=-1$).
Since the natural reflection symmetry operator for spin-$\frac12$
particles squares to $-1$ and commutes with $T$,
the reflection symmetry required for $\Z_2$ Weyl semimetals is
a special reflection symmetry,
which corresponds to a combination of natural reflection
and $\pi$ rotation in the spin space.

\begin{table}[tb]
\begin{center}
\caption{\label{table: topological charges}
Topological charge that is assigned to gap-closing points
in the three-dimensional Brillouin zone
under various symmetry constraints which are chosen from
time-reversal symmetry $T$,
reflection symmetry $R$,
and inversion symmetry $P$.
We assume that the gap closing does not take place
at time-reversal invariant momenta.
In cases where there are multiple symmetries,
the type of topological charge depends on the sign of squares of
the combined symmetry operator.
The reflection $R$ that gives $(TR)^2=-1$ actually means
combination of reflection and $\pi$ rotation in spin space
for spin-$\frac12$ electrons.
The case where $(TP)^2=+1$ can be realized in
semimetals with time-reversal, inversion, and
SU(2) spin rotation symmetries; see Sec.~IIID.
\\
}
\begin{tabular}[t]{cc}
\hline \hline
 ~Symmetry~ & ~Charge~ \\
\hline
 no symmetry & $\mathbb{Z}$ \\
 $T$ & $\mathbb{Z}$ \\
 $P$ & $\mathbb{Z}$ \\
 $T$ and $R$: $(TR)^2=+1$ & $0$ \\
 $T$ and $R$: $(TR)^2=-1$ & $\mathbb{Z}_2$ \\
 $T$ and $P$: $(TP)^2=-1$ & $0$ \\
 $T$ and $P$: $(TP)^2=+1$ & $\mathbb{Z}_2$ \\
\hline \hline
\end{tabular}
\end{center}
\end{table}

We note that $\Z_2$ Weyl/Dirac semimetals
are different from Dirac semimetals in which
Dirac points located at high symmetry points
in the Brillouin zone are protected
by crystalline symmetries\cite{Young-Dirac-semimetal12,Wang12,Wang13,steinberg14}
and which are recently reported\cite{Neupane13,Borisenko13,Liu14}
to be realized in Cd$_3$As$_2$ and Na$_3$Bi.
In contrast to these Dirac semimetals with nontrivial crystalline
symmetries, $\Z_2$ Weyl (Dirac) semimetals that we propose in this paper
have Weyl (Dirac) points with $\Z_2$ topological charge
which are stabilized by the interplay of
time-reversal symmetry and reflection (inversion) symmetry.

The plan of this paper is as follows.
In Sec.~II we introduce $\Z_2$ Weyl semimetals under
the presence of both time-reversal and reflection symmetries.
We present several toy models of $\Z_2$ Weyl semimetals
and show their energy spectra.
In Sec.~III 
we study the stability of these gap-closing points
for various cases by examining whether
the low-energy Dirac Hamiltonian can admit a Dirac mass term under
given symmetry constraints.
This task is accomplished by making use of Clifford algebras and
their representation theory.\cite{kitaev09,morimoto-clifford13}
We show that a pair of Weyl points are stable and have
$\Z_2$ topological charge under both time-reversal and reflection
symmetries.
We further show that a Dirac point with $\Z_2$ topological charge
is stabilized under SU(2) spin, time-reversal, and inversion symmetries.
The stability of point nodes with $\Z_2$ topological charge
in superconductors is also discussed.
In the Appendix we explain the basic idea of the stability analysis
using Clifford algebras and its application
to Dirac Hamiltonians in all the ten Altland-Zirnbauer symmetry classes.

\section{$\mathbb{Z}_2$ Weyl semimetals}

\subsection{Time-reversal and reflection symmetries}

In this section we discuss Weyl semimetals
with both time-reversal symmetry and 
reflection symmetry.
Time-reversal symmetry is represented by an anti-unitary operator,
while reflection symmetry is represented by a unitary operator $R_z$
with a mirror plane assumed to be perpendicular to the $z$ direction.
Under these symmetries, the three-dimensional Bloch Hamiltonian
satisfies the relations
\begin{subequations}
\label{eq: symmetry relations}
\begin{align}
TH(-k_x,-k_y,-k_z)T^{-1}&=H(k_x,k_y,k_z), \\
R_zH(k_x,k_y,-k_z)R_z^{-1}&=H(k_x,k_y,k_z).
\end{align}
\end{subequations}
Suppose that a Weyl point is located at $\bm{k}=(k_x^0,k_y^0,k_z^0)$
which is neither a high-symmetry point nor a time-reversal invariant momentum.
The time-reversal and reflection symmetries imply that
there are three other associated Weyl points:
$\bm{k}=(-k_x^0,-k_y^0,-k_z^0)$,
$(k_x^0,k_y^0,-k_z^0)$, and $(-k_x^0,-k_y^0,k_z^0)$.
Operations of $T$ and $R_z$ are not closed
for a single Weyl point but couple Weyl points (valleys).
Incidentally, if two Weyl points $(k_x^0,k_y^0,k_z^0)$ and
$(-k_x^0,-k_y^0,k_z)$ happen to be identical modulo reciprocal lattice
vectors, then the pair of Weyl points are combined to form
a Dirac point.
We will consider such a case in the next section.

Let us assume that the low-energy effective Hamiltonian
has translation symmetry and vanishing inter-valley coupling.
\footnote{Without these assumptions, gap-closing (Dirac or Weyl) points
can be always gapped out by inter-valley scattering.}
For the low-energy Hamiltonian of a pair of Weyl points
(or a single Dirac point) on the $k_z=k_z^0$ plane,
$T$ and $R_z$ are not symmetry operations, 
but the product $R_zT$ is.
We thus define the combined symmetry operator
\begin{align}
\widetilde T = R_zT,
\end{align}
which is an antiunitary operator satisfying
\begin{align}
\widetilde T H(-k_x,-k_y,k_z) \widetilde T^{-1}=H(k_x,k_y,k_z).
\label{eq: TRz symmetry}
\end{align}
The $\widetilde T$ operator relates a pair of Weyl points at,
e.g., $\bm{k}=(k_x^0,k_y^0,k_z^0)$ and $(-k_x^0,-k_y^0,k_z^0)$.

We now assume that
\begin{align}
\widetilde T^2=-1.
\label{eq: square TR}
\end{align}
As we show below, Eq.\ (\ref{eq: square TR}) is
the essential condition for the existence of $\Z_2$ Weyl semimetals.%
\footnote{
For systems of interacting fermions,
the right-hand side of Eq.~(\ref{eq: square TR}) is
replaced with fermion number parity, $(-1)^{N_F}$,
where $N_F$ is the number of fermions.
}
Some comments on reflection (mirror) symmetry are in order here.
For spin-1/2 fermions, time-reversal transformation
takes the form $T=i\sigma_y\mathcal{K}$, where $\mathcal{K}$ is
a complex conjugation operator.
Reflection with respect to a mirror plane ($z=0$, say) involves
$\pi$ rotation of spin and is given by $R_z=i\sigma_z$,
which leads to $\widetilde T^2=+1$.
However, we can consider cases when 
the Hamiltonian is invariant under $R_z=1$ 
(i.e., without $\pi$ spin rotation),
which results in $\widetilde T^2=-1$.
Some model Hamiltonians with $\widetilde T^2=-1$ will be discussed
in the next section.

\begin{figure}
\begin{center}
\includegraphics[width=\linewidth]{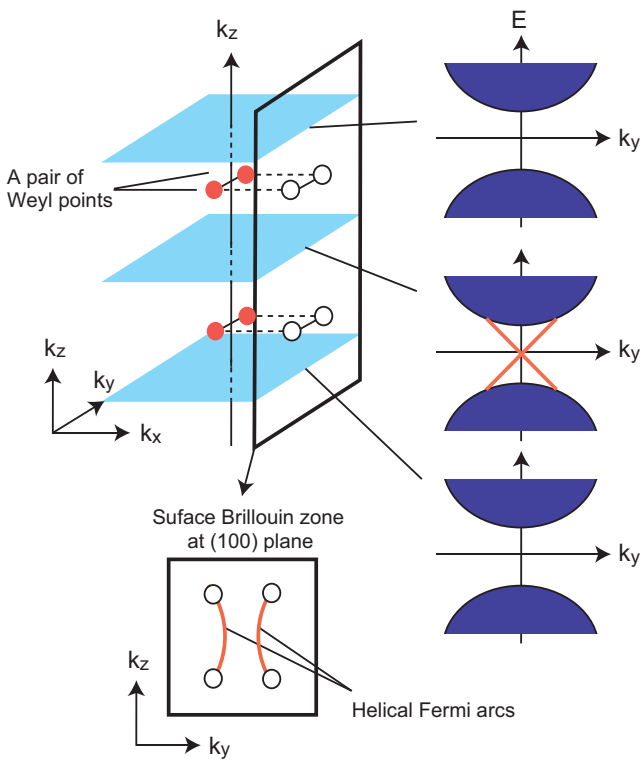}
\end{center}
\caption{
Schematic picture of a $\mathbb{Z}_2$ Weyl semimetal.
Helical Fermi arcs appear between a time-reversal pair of
$\mathbb{Z}_2$ Weyl points.
A surface perpendicular to the $x$ direction
has helical edge states in the surface band structure 
as a function of $k_y$ with fixed $k_z$
between two pairs of Weyl points, as depicted in the right panel.
}
\label{Fig: Fermi arc}
\end{figure}

Equations (\ref{eq: TRz symmetry}) and
(\ref{eq: square TR}) imply that, for each fixed value of $k_z$,
$H(k_x,k_y,k_z)$ can be regarded as a Hamiltonian that is
invariant under $\widetilde T$ in the 2D Brillouin zone $(k_x,k_y)$.
This means that $H$ is effectively a 2D Hamiltonian of class AII
in the Altland-Zirnbauer classification
of free-fermion Hamiltonians.\cite{AZ-classes}
Consequently,
for any 2D plane of fixed $k_z$ on which $H(k_x,k_y,k_z)$ is gapped,
we can define the $\mathbb{Z}_2$ topological index $\nu_2(k_z)$,
in the same way as in the 2D $\Z_2$ topological
insulators;\cite{kane2005z_,fu-kane-pump06}
\begin{align}
(-1)^{\nu_2(k_z)}=\prod_{(k_x,k_y) \in \t{TRIM}_2}
\frac{\t{Pf}\,[w(k_x,k_y;k_z)]}{\sqrt{\t{det}[w(k_x,k_y;k_z)]}}
\label{eq: Z2 for T+R}
\end{align}
with
\begin{align}
w_{ij}(k_x,k_y;k_z)&= 
\langle \psi_i(-k_x,-k_y,k_z)|\widetilde T| \psi_j(k_x,k_y,k_z) \rangle,
\end{align}
where TRIM$_2$ denotes momenta which are invariant under the action of 
time-reversal transformation
on the 2D plane of constant $k_z$,  
and $|\psi_i(k_x,k_y,k_z) \rangle$ is 
a wave function of the $i$th valence band defined smoothly over the whole
plane of $(k_x,k_y)$.
The $\mathbb{Z}_2$ topological index $\nu_2(k_z)$ can change only
when $k_z$ is varied across the plane containing a pair of Weyl points.
This change in $\nu_2(k_z)$ is assigned to the pair of Weyl points
as $\mathbb{Z}_2$ topological charge.
Suppose that a $k_x$-$k_y$ plane between two pairs of Weyl points
has $\nu_2(k_z)=1$, as shown in Fig.~\ref{Fig: Fermi arc}.
In this case the surface Brillouin zone
$(k_y,k_z)$ of a (100) surface has
a pair of Fermi arcs (helical Fermi arcs)
coming from helical surface states whose existence is guaranteed by
$\nu_2(k_z)=1$, as schematically shown in Fig.~\ref{Fig: Fermi arc}.
The helical Fermi arcs connect Weyl points projected
onto the surface Brillouin zone.

These features clearly indicate that $\Z_2$ Weyl semimetals
are time-reversal invariant $\Z_2$ version of conventional
Weyl semimetals in which Weyl points have integer topological
charges and Fermi arcs are formed by chiral surface states.

Finally, we emphasize that the topological stability of a pair of Weyl points
on a 2D plane of constant $k_z$ come from the assumed conditions
of the $R_z T$ symmetry,
Eqs.\ (\ref{eq: TRz symmetry}) and (\ref{eq: square TR}).
In fact, two Weyl points forming a $\widetilde T$-invariant pair
in a $\Z_2$ Weyl semimetal
are a source and a drain of Berry curvature
and can be assigned integer topological charges of opposite signs.
Since a pair of Weyl points are charge neutral as a whole,
they could merge and pair-annihilate.
However, with the conditions in
Eqs.\ (\ref{eq: TRz symmetry}) and (\ref{eq: square TR}),
a $\Z_2$ topological charge is given to a pair of Weyl points
as a whole, which prohibits pair-annihilation
even when they merge at a TRIM$_2$.

\subsection{Examples}

In this section we present four tight-binding models of
$\Z_2$ Weyl semimetals.
In these models the condition of Eq.\ (\ref{eq: square TR}) is implemented by
$\widetilde T=R_zT$ with
\begin{align}
T^2=-1,
\quad
R_z^2=1, \quad [T,R_z]=0.
\label{eq: T^2=-1}
\end{align}
In all the following models we set the Fermi velocity to be 1.

The first example is a 3D variant of the Bernevig-Hughes-Zhang (BHZ) model 
and is given by the Bloch Hamiltonian 
\begin{align}
H_1=&\,{}
\tau_x (\sigma_z \sin k_y + v) + \tau_y \sin k_x \n
& + \tau_z (M-\cos k_x -\cos k_y - \cos k_z).
\label{eq: stacked BHZ model}
\end{align}
Here $\sigma_\alpha$ and $\tau_\alpha$ are Pauli matrices corresponding to
spin and orbital degrees of freedom.
For $v=0$ and fixed $k_z$,
$H_1$ has the same form as the BHZ model,\cite{Bernevig15122006}
and indeed $H_1$ is obtained by stacking the 2D BHZ model
along the $z$ direction.
The Hamiltonian satisfies the symmetry relations of
Eqs.\ (\ref{eq: symmetry relations}) with
\begin{equation}
T=i\sigma_y \mathcal{K},
\qquad
R_z=1. 
\label{eq: symmetry T=i sy K and R_z=1}
\end{equation}
When $v=0$ and $M=2$,
we have two Dirac points at
\begin{align}
\bm{k}=(0,0,\pm\pi/2)
\end{align}
in the Brillouin zone $-\pi\le k_\alpha\le \pi$.
The $\mathbb{Z}_2$ topological number $\nu_2(k_z)$ is obtained
as a function of $k_z$ from Bloch wave functions of $H_1$;
\begin{equation}
\nu_2(k_z)=\left\{\begin{array}{ll}
0, & -\pi\le k_z<-\pi/2,\\
1, & -\pi/2<k_z<\pi/2,\\
0, & \pi/2<k_z\le\pi.
\end{array}\right.
\label{eq: range of nu_2(k_z)}
\end{equation}
The two Dirac points separate the regions of different values of $\nu_2(k_z)$.
When the parameter $v$ is finite,
each Dirac point splits into two Weyl points 
which are on the same $k_z$ plane 
(that is slightly shifted from $k_z=\pm \pi/2$)
and are related to each other by $\widetilde T$.

The second example is a stacked Kane-Mele model
defined on the stacked layers of the honeycomb lattice.
The Hamiltonian for an electron with spin $\bm{s}$ and wave number $k_z$
along the stacking direction is given by
\begin{align}
{\cal H}_2=&\,{}
t\sum_{\langle i,j \rangle}c_i^\dagger c_j +
i(\lambda_{SO}+\lambda_{SO}' \cos k_z) 
\sum_{\langle\langle i,j \rangle\rangle}\nu_{ij}c_i^\dagger s_z c_j \n
& +
i \lambda_{R}\sum_{\langle i,j \rangle}
c_i^\dagger(\bm{s} \times \Vec d_{ij})_z c_j +
\lambda_{v}\sum_i \xi_i c_i^\dagger c_i,
\label{eq: stacked Kane-Mele model}
\end{align}
where we have followed the standard notation used in the
Kane-Mele model.\cite{kane-mele05,kane2005z_}
The first term is a nearest-neighbor hopping term on the honeycomb lattice,
where $c_j=(c_{j,\uparrow}, c_{j,\downarrow})$ annihilates an electron on site $j$.
The second term is a spin-dependent second-neighbor hopping term with
$\nu_{ij}=(2/\sqrt{3})(\Vec d_1 \times \Vec d_2)_z=\pm1$,
where $\Vec d_1$ and $\Vec d_2$ are unit vectors along the two bonds which
an electron traverses when going from site $j$ to $i$.
We have included a small spin-dependent hopping
between neighboring layers with amplitude $\lambda_{SO}'$.
We assume that  the interlayer coupling is present only in this form.
The third term is a nearest-neighbor Rashba term induced by breaking of
inversion along the $z$ direction.
The vector $\Vec d_{ij}$ is a unit vector pointing from site $j$ to $i$.
The last term is the staggered potential with $\xi=+1$ for one sublattice
and $\xi=-1$ for the other sublattice of the honeycomb lattice.
With $\lambda_{SO}'=0$, the above Hamiltonian $\mathcal{H}_2$ in
Eq.~(\ref{eq: stacked Kane-Mele model}) is in the same form as
the Kane-Mele model.\cite{kane-mele05,kane2005z_}
The Hamiltonian $\mathcal{H}_2$ satisfies the time-reversal and
reflection symmetry relations in
Eq.~(\ref{eq: symmetry T=i sy K and R_z=1}).

The third example 
is given by a Bloch Hamiltonian
on the cubic lattice:
\begin{align}
H_3=&\,{}
\sigma_x \tau_z \sin k_x + \sigma_y \tau_z \sin k_y \n
& + \tau_x (\cos k_x +\cos k_y +\cos k_z  -M).
\label{eq: H Dirac semimetal}
\end{align}
Here $\sigma_\alpha$ and $\tau_\alpha$ are Pauli matrices corresponding to
spin and orbital degrees of freedom.
The first two terms in Eq.\ (\ref{eq: H Dirac semimetal}) represent
spin-orbit coupling of the Rashba type, with opposite signs for
the two orbitals labeled by $\tau_z=\pm 1$.
The third term represents hopping between different
orbitals on nearest-neighbor sites.
The Hamiltonian satisfies the symmetry relations of
Eqs.\ (\ref{eq: symmetry relations}) with the symmetry operators
given in Eq.\ (\ref{eq: symmetry T=i sy K and R_z=1}).
When we set $M=2$,
we have two Dirac points at
$
\bm{k}=(0,0,\pm\pi/2)
$
and the $\mathbb{Z}_2$ topological number $\nu_2(k_z)$ given by
Eq.~(\ref{eq: range of nu_2(k_z)}).

The last example is also given by a  Hamiltonian
defined on the cubic lattice:
\begin{align}
H_4=&\,{}
\tau_x \cos k_y + \tau_y \sin k_x + \tau_z \sigma_z \sin k_y \n
& + \tau_z (2-\cos k_x -\cos k_z).
\label{eq: H for square lattice 2}
\end{align}
Again the Hamiltonian is invariant under time-reversal transformation
and reflection defined
by Eq.~(\ref{eq: symmetry T=i sy K and R_z=1}).

\begin{figure*}
\begin{center}
\includegraphics[width=0.8\linewidth]{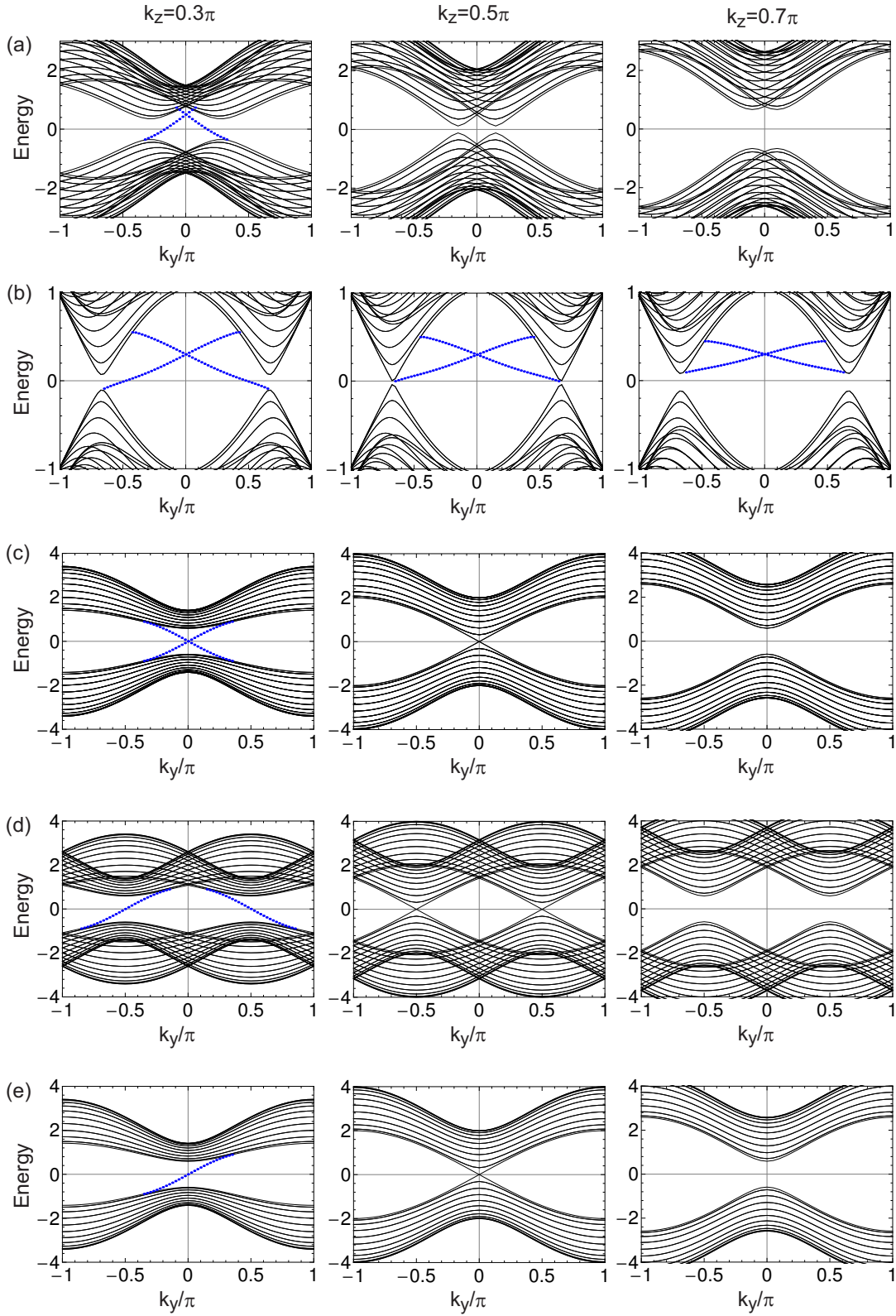}
\end{center}
\caption{
Band structures of tight-binding models for $\Z_2$ Weyl semimetals (a)-(d)
and a Weyl semimetal (e):
(a) stacked BHZ model $H_1$ [Eq.~(\ref{eq: stacked BHZ model}) with $v=0.5$],
(b) stacked Kane-Mele model $\mathcal{H}_2$
 [Eq.~(\ref{eq: stacked Kane-Mele model})
with $(t,\lambda_{SO},\lambda_{SO}',\lambda_R,\lambda_v)=(1,0.06,0.03,0.05,0.3)$],
(c) $H_3$ [Eq.~(\ref{eq: H Dirac semimetal})],
(d) $H_4$ [Eq.~(\ref{eq: H for square lattice 2})],
(e) Weyl semimetal $H_5$ [Eq.~(\ref{eq: H Weyl semimetal})].
These models have a two-dimensional (100) surface which is perpendicular
to the $x$ direction;
the 2D surface of the stacked Kane-Mele model (b) is coupled
zigzag edges running along the $y$ direction.
Periodic boundary conditions are assumed in the $y$ and $z$ directions.
Band structures are shown as functions of $k_y$ for fixed
$k_z=0.3 \pi, 0.5 \pi, 0.7\pi$.
Black lines are bulk bands while blue dots are
surface modes localized at the (100) surface.
Surface states of other surfaces are not shown.
The upper and lower bands of the models (c)-(e) touch when $k_z=\pi/2$.
}
\label{Fig: helical edge}
\end{figure*}

In Fig.~\ref{Fig: helical edge}(a)-(d)
we show the bulk and surface band structure of the models defined in
Eqs.\ (\ref{eq: stacked BHZ model}), (\ref{eq: stacked Kane-Mele model}),
(\ref{eq: H Dirac semimetal}), and (\ref{eq: H for square lattice 2}).
For comparison,
we also show in Fig.~\ref{Fig: helical edge}(e)
the bulk and surface band structure of a model for a conventional
Weyl semimetal described 
by the Hamiltonian
\begin{align}
H_W=&\;{}
\sigma_x \sin k_x + \sigma_y \sin k_y \n
& + \sigma_z (\cos k_x +\cos k_y +\cos k_z  - M),
\label{eq: H Weyl semimetal}
\end{align}
where we set $M=2$ to have Weyl points at $\Vec k=(0,0,\pm\pi/2)$.
The energy spectra of these tight-binding models
(except the stacked Kane-Mele model) are
studied for the cubic lattice with a (100) surface.
The stacked Kane-Mele model $\mathcal{H}_2$
(\ref{eq: stacked Kane-Mele model}) is solved for a lattice
obtained by stacking (in the $z$ direction) layers of the honeycomb lattice
with a zigzag edge running along the $y$ direction.
In solving these models numerically, we have assumed
periodic boundary conditions in the $y$ and $z$ directions
and open 
boundary conditions in the $x$ direction
(i.e., vanishing matrix elements for hopping out of the surface).

In Fig.~\ref{Fig: helical edge}, the energy bands
are plotted as functions of $k_y$
for fixed values of $k_z$, $k_z=0.3 \pi,0.5\pi,0.7\pi$.
In the figures solid black lines are bulk bands and
blue dots are surface states localized near one surface
perpendicular to the $x$ axis
(surface states localized near other surfaces are not shown in the figures).
Figure~\ref{Fig: helical edge} clearly shows that,
at $k_z=0.3\pi$, the $\Z_2$ Weyl semimetals
have helical modes while the Weyl semimetal has a chiral mode.
These modes form Fermi arcs in the surface Brillouin zone.
As $k_z$ is increased,
the band gap closes at $k_z=0.5\pi$ in Fig.~\ref{Fig: helical edge}(c)-(e)
[$k_z\approx0.5\pi$ in Fig.~\ref{Fig: helical edge}(a), (b)].
When the band gap reopens ($k_z>0.5\pi$), surface modes connecting
the upper and lower bands disappear, as seen in the figures
for $k_z=0.7\pi$.

We note that the Hamiltonian $H_3$
in Eq.~(\ref{eq: H Dirac semimetal})
has additional particle-hole symmetry
$C=\sigma_x \tau_z \cal K$ and
unitary symmetry $U=\sigma_z \tau_x$.
Indeed, if we exchange $\tau_x$ and $\tau_z$
in Eq.\ (\ref{eq: H Dirac semimetal}), $H_3$ becomes
a Bogoliubov-de Gennes Hamiltonian of the planar state of
a $p$-wave superconductor,
which has time-reversal, particle-hole, and $U(1)$ spin rotation symmetries
($S_z$ conservation).\cite{vollhardt2003superfluid,makhlin-planar13}
The planar state also has point nodes and surface modes counter-propagating
for opposite spins,
but it is characterized by an integer topological number
rather than a $\Z_2$ topological number.\cite{makhlin-planar13}
In the basis where $U$ is diagonalized,
we can define a Chern number for each spin sector for a fixed value of $k_z$.
In this sense the planar state is considered as two copies of
the $^3$He-A phase\cite{volovik2003universe}
which has a chiral surface mode and a Fermi arc.
However, in our example of the $\mathbb{Z}_2$ Weyl semimetal of 
Eq.~(\ref{eq: H Dirac semimetal}),
we can break the particle-hole symmetry and the unitary symmetry
by adding perturbations
which keep the time-reversal and reflection symmetries intact
(such as $\sigma_y\tau_y$, $\sigma_z\tau_y$, and $\sigma_z\tau_z\sin k_x$).
The breaking of the particle-hole and unitary symmetries does not
affect the $\mathbb{Z}_2$ topological index 
in Eq.~(\ref{eq: Z2 for T+R}).
Therefore the essential symmetry for stabilizing $\Z_2$ Weyl
semimetals is the product symmetry $\widetilde T$ with $\widetilde T^2=-1$.
The realization of this symmetry is not limited to the one we
discussed above, Eq.\ (\ref{eq: T^2=-1}).
For example, another way to realize the combined symmetry $\widetilde T^2=-1$
would be
\begin{align}
T^2=+1, \quad
R_z^2=1, \quad \{T,R_z\}=0.
\end{align}

\section{Stability analysis of Weyl and Dirac points using Clifford algebras}

In this section we discuss stability of gap-closing (Weyl or Dirac) points
in semimetals without/with time-reversal symmetry and other symmetries,
and we further determine the type of topological charge attached to
gap-closing points.
In fact, the stability of Fermi points has been previously studied
by applying K-theory.\cite{horava05,zhao-wang13,matsuura13,shiozaki14}
Here we study the stability of Weyl/Dirac points by examining whether
the effective theory for excitations near a gap-closing point can have a
Dirac mass term compatible with symmetry constraints.
For this purpose, we use representation theory of Clifford algebras
and K-theory.\cite{kitaev09,morimoto-clifford13}
In the Appendix we explain this approach (i.e., existence condition
of a Dirac mass term) and apply it to all
the ten Altland-Zirnbauer symmetry classes.\cite{AZ-classes}
Below we apply the approach to the cases with spatial symmetries
to find types of topological charges that emerge
under a given set of symmetries (Table~\ref{table: topological charges}).

\subsection{Weyl semimetal}
As is well known, a Weyl point is stable and has an integer topological charge
in three dimensions, when low-energy effective theory of the Weyl point has
no symmetry.\cite{Wan-semimetal11,burkov-balents11,murakami-semimetal07,Yang-Lu-Ran11,volovik2003universe}
We will derive this known fact using representation theory
of complex Clifford algebras,
as a prelude to the stability analysis under
time-reversal symmetry which we will present in the following subsections.

A complex Clifford algebra $Cl_{q}$ is a complex algebra
generated by $q$ generators ($e_1,\ldots,e_{q}$) satisfying
\begin{align}
\{e_i^{},e_j^{}\}&= 2\delta_{i,j},
\end{align}
with Kronecker's $\delta_{i,j}$.
In this paper we use the notation
\begin{equation}
Cl_{q}=\{e_1,\ldots,e_{q}\}
\end{equation}
to represent the whole complex algebra $Cl_{q}$
generated from the $q$ generators ($e_1,\ldots,e_{q}$).

As an effective Hamiltonian for low-energy excitations
around a Weyl point,
we consider a three-dimensional Dirac (Weyl) Hamiltonian
\begin{align}
H_\mathrm{eff}=k_x \gamma_x+k_y \gamma_y+k_z \gamma_z + m \gamma_0,
\label{eq: 3D Weyl H}
\end{align}
where $\gamma_j$ ($j=0,x,y,z$) are gamma matrices satisfying
the anticommutation relations $\{\gamma_j,\gamma_l\}=2\delta_{j,l}$.
We assume that $(k_x,k_y,k_z)$ are momenta 
measured from the Weyl point and that
the Weyl point is not located at a high symmetric point.
We have included a Dirac mass term
$m\gamma_0$ in Eq.\ (\ref{eq: 3D Weyl H}) 
which would gap out the Weyl point.
We will examine whether such a mass term is allowed
when kinetic terms are given.
If it is not allowed, then
the Weyl point is stable against (translation-invariant) perturbations.

The Hamiltonian of a single Weyl point (\ref{eq: 3D Weyl H})
has no symmetry and is classified as a member of class A.
In this case a complex Clifford algebra is generated by the gamma matrices
in the Dirac Hamiltonian as
\begin{align}
Cl_4=\{\gamma_x,\gamma_y,\gamma_z,\gamma_0 \}.
\end{align}
The answer to the question as to whether a mass term $\gamma_0$ is allowed
is obtained by studying the topological classification
of a generator (say, $\gamma_z$)
of the Clifford algebra without $\gamma_0$,
\begin{equation}
Cl_3=\{\gamma_x,\gamma_y,\gamma_z \}.
\end{equation}
This is because
topologically trivial classification of $\gamma_z$
implies the existence of another gamma matrix (i.e., $\gamma_0$)
which anticommutes with the three generators
($\gamma_x$, $\gamma_y$, and $\gamma_z$),
while
the topologically nontrivial classification of $\gamma_z$ implies
the absence of $\gamma_0$;
see Appendix.

We thus consider the following extension problem of Clifford algebra,
\begin{align}
Cl_{2}=\{\gamma_x,\gamma_y\} &\to 
Cl_{3}=\{\gamma_x,\gamma_y,\gamma_z \}.
\label{eq: Cl_2 to Cl_3}
\end{align}
We first fix a matrix representation (of sufficiently large dimensions)
of the original algebra $Cl_2$ and ask how many distinct classes
of matrix representations we have for the added generator ($\gamma_z$)
in the extended algebra $Cl_3$.
It is known from K theory
that all the possible matrix representations form a symmetric space,
i.e., classifying space.\cite{kitaev09}
The classifying space for the extension problem (\ref{eq: Cl_2 to Cl_3})
is known to be $C_0=\cup_{m \in \mathbb{Z}} U(2n)/[U(n+m) \times U(n-m)]$ 
with a sufficiently large integer $n$, i.e.,
a union of complex Grassmanians;
see, for more details,
Refs.~\onlinecite{kitaev09} and \onlinecite{morimoto-clifford13}.
Its zero-th homotopy group,
\begin{align}
\pi_0(C_0)&=\mathbb{Z},
\label{eq: pi0 of C_0}
\end{align}
indicates that the space of all possible representations of $\gamma_z$
consists of disconnected parts, which can be labelled
with an integer topological index.
The nontrivial topology of the space of $\gamma_z$ also means that
a Dirac mass term is not allowed in the minimal Dirac Hamiltonian
(\ref{eq: 3D Weyl H}).
Hence a Weyl point is stable against (spatially uniform) perturbations.
The integer topological index corresponds to
the Chern number of a 2D subsystem with fixed $k_z$ in which
$k_z \gamma_z$ behaves as a mass term
(the sign of $k_z$ is related to the Chern number).
With $k_z$ taken as a tuning parameter in the effective Hamiltonian,
the Weyl point can be viewed as a quantum phase transition point of
the 2D subsystem and is characterized by a $\mathbb{Z}$ charge
which is equal to the change in the Chern number.

An example of Weyl points is point nodes at the north and south poles
$\bm{k}=(0,0,\pm k_F)$ on the Fermi surface in the superfluid
$^3$He-A phase.
Each of the two point nodes is a
Weyl point described by an effective 2 by 2
Hamiltonian.\cite{volovik2003universe,silaev12}
Stability of point nodes in $^3$He-A with particle-hole symmetry
is understood using Clifford algebras as follows.
The particle-hole symmetry is described by an anti-unitary
operator $C=\tau_x \cal K$,
where $\tau_x$ is a Pauli matrix acting on the particle-hole space.
However, action of $C$ connects two point nodes at $\Vec k=(0,0,\pm k_F)$ 
and is not closed for a single point node (Weyl point).
Hence the Bogoliubov-de Gennes Hamiltonian for quasiparticles of
a single point node
has no symmetry and is classified into class A.
Thus the stability of Weyl point nodes can be explained
in the same manner as described above.

In the presence of additional spatial symmetries,
topological characterization of gap-closing points
in superconductors may change,
as we discuss for Weyl/Dirac semimetals in the following subsections.
We note that stability of line
nodes\cite{sato06,beri10,schnyder-ryu11}
was recently studied for superconductors with inversion symmetry or
reflection symmetry
and for odd-parity superconductors in
Ref.~\onlinecite{kobayashi14}.
Study of stable point nodes accompanied by nontrivial surface states\cite{volovik2003universe,schnyder-ryu11,heikkila11,tsutsumi10,wang-lee12}
has been expanded to include cases with
reflection symmetry\cite{zhang-kane-tmsc,Yao-Ryu13} and those with
reflection and inversion symmetries.\cite{yang14}
Two nontrivial examples of point nodes in topological superconductors
will be discussed in Sec.~\ref{sec: Z2 nodes}.

\subsection{Time-reversal and reflection symmetries:\\ $\Z_2$ Weyl semimetal}
In this section we show stability
of Weyl points with $\mathbb{Z}_2$ charge
under time-reversal and reflection symmetries
using Clifford algebras.
As we discussed in Sec.~IIA, in the presence of the two symmetries,
we have a quartet of Weyl points at
$\bm{k}=(k_x^0,k_y^0,k_z^0)$, $(-k_x^0,-k_y^0,k_z^0)$,
$(k_x^0,k_y^0,-k_z^0)$, and $(-k_x^0,-k_y^0,-k_z^0)$.
Since a pair of Weyl points $(k_x^0,k_y^0,k_z^0)$ and $(-k_x^0,-k_y^0,k_z^0)$
are related by the combined symmetry $\widetilde T=R_zT$,
we treat them together as a single Dirac point
and set $k_x^0=k_y^0=0$ to simplify notation.
Incidentally, this also accounts for the special case where
$(k_x^0,k_y^0)\in\mathrm{TRIM}_2$, as in the case shown
in Fig.~\ref{Fig: helical edge}(c).

As an effective Hamiltonian for low-energy excitations
around the Dirac point,
we consider a three-dimensional Dirac Hamiltonian
\begin{align}
\widetilde{H}_\mathrm{eff}
=k_x \gamma_x+k_y \gamma_y+(k_z - k_z^0) \gamma_z + m \gamma_0,
\label{eq: 3D Dirac H}
\end{align}
where $\gamma_j$ ($j=0,x,y,z$) are gamma matrices satisfying
the anticommutation relations $\{\gamma_j,\gamma_l\}=2\delta_{j,l}$.
We assume that the Dirac point $(0,0,k_z^0)$ and its time-reversal
partner $\bm{k}=(0,0,-k_z^0)$ are distinct points in the Brillouin zone.
In the following discussions we consider only
low-energy excitations around the Dirac point at $\bm{k}=(0,0,k_z^0)$,
because we are concerned with the stability of individual Dirac points
against translation-invariant perturbations.
As an example of such a perturbation,
we have included a Dirac mass term
$m\gamma_0$ in Eq.\ (\ref{eq: 3D Dirac H}) 
which would gap out the Dirac point.
We will examine whether this mass term
is compatible with the assumed symmetries.
If it is not compatible, then
the Dirac point is stable against (translation-invariant) perturbations.

Since $T$ or $R_z$ alone is not a symmetry of the effective
Hamiltonian $\widetilde{H}_\mathrm{eff}$,
the only symmetry operator for $\widetilde{H}_\mathrm{eff}$
is the product $\widetilde T=R_zT$,
which is assumed to satisfy $\widetilde T^2=-1$.
Whether or not a Dirac mass term can exist under this symmetry
is systematically studied using Clifford
algebras below.\cite{kitaev09,morimoto-clifford13,morimoto-weakTI14}
From Eq.\ (\ref{eq: TRz symmetry}) we find that
the $\widetilde T$ symmetry and gamma matrices satisfy
the following algebraic relations:
\begin{subequations}
\label{eq: TR_z and gamma}
\begin{align}
\{\gamma_x,\widetilde T\}=\{\gamma_y,\widetilde T\}&=0, \\
[\gamma_z,\widetilde T]=[\gamma_0,\widetilde T]&=0. 
\end{align}
\end{subequations}
We treat the symmetry operator $\widetilde T$ and the gamma matrices $\gamma_i$
on equal footing in real Clifford algebras.
A real Clifford algebra $Cl_{p,q}$ is a real algebra
generated by $p+q$ generators ($e_1,\ldots,e_{p+q}$) satisfying
\begin{subequations}
\begin{align}
\{e_j^{},e_l^{}\}&= 0 ~~ (j\neq l), \\
e_j^2 &=
\left\{
\begin{array}{ll}
-1, ~~ 1\le j \le p, \\
+1, ~~ p+1\le j \le p+q.
\end{array}\right.
\end{align}
\end{subequations}
In this paper we use the notation
\begin{equation}
Cl_{p,q}=\{e_1,\ldots,e_p; e_{p+1},\ldots,e_{p+q}\}
\end{equation}
to represent the whole real algebra $Cl_{p,q}$
generated from the $p+q$ generators ($e_1,\ldots,e_{p+q}$).
To incorporate the antiunitary nature of the $\widetilde T$ operator
in real algebras, we introduce an operator $J$ which plays a role
of the imaginary unit $i$ and anticommutes with $\widetilde T$,
\begin{equation}
J^2=-1,
\qquad
\{\widetilde T, J\}=0.
\label{eq: J}
\end{equation}
The gamma matrices commute with $J$, $[\gamma_i,J]=0$.

Using the symmetry relations in Eqs.~(\ref{eq: TR_z and gamma})
and (\ref{eq: J}),
we define the real Clifford algebra generated
from gamma matrices and the symmetry operator as
\begin{align}
Cl_{0,4} \tensor Cl_{0,2}=
\{\,;\gamma_x,\gamma_y,\gamma_z,\gamma_0\}\tensor
\{\,; \gamma_x \gamma_y \widetilde T, J\gamma_x \gamma_y \widetilde T \}.
\end{align}

From the argument explained in the Appendix,
the question as to whether a mass term $\gamma_0$ is allowed
under given symmetry is answered by considering
the classification problem of a generator of the same type 
as $\gamma_0$ (e.g., $\gamma_z$) for the Clifford algebra without $\gamma_0$,
\begin{equation}
Cl_{0,3} \tensor Cl_{0,2}=
\{\,;\gamma_x,\gamma_y,\gamma_z\}\tensor
\{\,; \gamma_x \gamma_y \widetilde T, J\gamma_x \gamma_y \widetilde T \}.
\label{eq: Cl0,3*Cl0,2}
\end{equation}
As in the discussion in Sec.~IIIA,
if the space of matrix representations of $\gamma_z$ is topologically trivial,
then there is another gamma matrix that can be used as $\gamma_0$.
On the other hand, if it is topologically nontrivial, then
there is no such gamma matrix, hence no $\gamma_0$.

We thus consider the extension problem of Clifford algebra
\begin{align}
Cl_{0,2} \tensor Cl_{0,2} &\to Cl_{0,3} \tensor Cl_{0,2}.
\label{extension Z_2}
\end{align}
We fix a matrix representation (in sufficiently large dimensions)
of $Cl_{0,2}\tensor Cl_{0,2}$ and ask how many possible matrix
representations we have for $\gamma_z$ in $Cl_{0,3}\tensor Cl_{0,2}$.
It turns out%
\footnote{
The extension problem in Eq.\ (\ref{extension Z_2}) is reduced to 
$Cl_{0,2}\to Cl_{0,3}$ as follows.
First, $Cl_{0,2}$ is isomorphic to the algebra of real 2 by 2 matrices $\mathbb{R}(2)$.
That is, if we take the representation $Cl_{0,2}=\{;\sigma_x,\sigma_z\}$,
an element of $Cl_{0,2}$ is written as
$a+b\sigma_x+ic\sigma_y+d\sigma_z$ with $a,b,c,d\in\mathbb{R}$,
which is a general form of a real 2 by 2 matrix.
Thus the extension problem in Eq.\ (\ref{extension Z_2}) is equivalent to
$Cl_{0,2}\tensor \mathbb{R}(2) \to Cl_{0,3}\tensor \mathbb{R}(2)$.
Next, $\mathbb{R}(2)$ can be discarded from the above extension,
because faithful representations of $A\tensor \mathbb{R}(n)$ in real matrices
have a natural one-to-one correspondence with those of $A$.%
}
that the space of representations
for $\gamma_z$ is given by 
the classifying space $R_2=O(2n)/U(n)$,
where $n$ is a sufficiently large integer and $2n$ is the dimension of 
representation.\cite{kitaev09,morimoto-clifford13}
Its zero-th homotopy group is known to be
\begin{align}
\pi_0(R_2)&=\mathbb{Z}_2.
\label{eq: pi0 of R_2}
\end{align}
This indicates that there is no mass term in the minimal
(4 by 4) Dirac Hamiltonian (or two 2 by 2 Weyl Hamiltonians),
while we can always find a mass term to gap out the Dirac point
if we double the minimal model.
This is precisely the $\mathbb{Z}_2$ nature of a pair of Weyl points.
Thus $\Z_2$ semimetal protected by time-reversal and reflection symmetries
with $\widetilde T^2=-1$ is characterized by $\mathbb{Z}_2$ charge
of a pair of Weyl points (or a Dirac point).

\subsection{Time-reversal and inversion symmetries}

As discussed in
Refs.~\onlinecite{murakami-semimetal07,Wan-semimetal11,burkov-balents11},
gap-closing points in a semimetal are fragile when
Hamiltonian is invariant under both time-reversal $T$
and inversion $P$.
Here we derive this known result using real Clifford algebras.

We consider a gap-closing (Weyl or Dirac) point at a generic $\bm{k}$ point
(not at one of time-reversal invariant momenta) in the Brillouin zone.
Separate operation of either time-reversal $T$ or inversion $P$
maps a Weyl/Dirac point at $\bm{k}=\bm{k}^0$ to another Weyl/Dirac point
at $\bm{k}=-\bm{k}^0$.
While neither time-reversal $T$ nor inversion $P$ is a closed operation
by itself,
the combination of the two operations $PT$ leaves
the effective Hamiltonian of a single Weyl/Dirac point
at $\bm{k}=\bm{k}^0$ invariant,
\begin{align}
PTH_\mathrm{eff}(k_x,k_y,k_z)(PT)^{-1}=H_\mathrm{eff}(k_x,k_y,k_z).
\label{eq: PT symmetry}
\end{align}
Substituting the Dirac Hamiltonian (\ref{eq: 3D Dirac H})
into the above equation,
we obtain symmetry relations obeyed by the gamma matrices,
\begin{align}
[\gamma_x,PT]=[\gamma_y,PT]=[\gamma_z,PT]=[\gamma_0,PT]=0.
\end{align}

Let us consider semimetals with strong spin-orbit coupling and inversion
symmetry.
We assume that the time-reversal operator $T$ and inversion operator $P$
satisfy the following relations:
\begin{align}
T^2=-1, \qquad
P^2=1, \qquad [T,P]=0,
\end{align}
thereby the combined operator $PT$ satisfying
\begin{align}
(PT)^2=-1.
\label{(PT)^2=-1}
\end{align}
We define a real Clifford algebra generated from $PT$ and 
gamma matrices,
\begin{align}
Cl_{0,4} \tensor Cl_{2,0}&=
\{ ;\gamma_x, \gamma_y, \gamma_z, \gamma_0 \} \tensor \{ PT, JPT ; \}.
\end{align}
The existence/absence of the Dirac mass $m\gamma_0$ can be judged
by considering the following extension problem:
\begin{align}
\{ ;\gamma_x, \gamma_y \} \tensor \{ PT, JPT ; \}
\to \{ ;\gamma_x, \gamma_y, \gamma_z \} \tensor \{ PT, JPT ; \},
\end{align}
i.e.,
\begin{align}
Cl_{0,2} \tensor Cl_{2,0} \to Cl_{0,3} \tensor Cl_{2,0},
\label{eq: extension R6}
\end{align}
which is equivalent to $Cl_{0,6}\to Cl_{0,7}$.%
\footnote{We take
the tensor product of $Cl_{0,2}$ and each side of Eq.~(\ref{eq: extension R6})
and make use of the following relations:
\begin{align*}
Cl_{p,q}\tensor Cl_{2,0}\simeq  Cl_{q+2,p},
\end{align*}
i.e.,
\begin{align*}
&\{e_1,\ldots,e_p;e_{p+1},\ldots,e_{p+q}\}\tensor\{\tilde e_1, \tilde e_2;\} \\
&\simeq \{\tilde e_1, \tilde e_2, \tilde e_1 \tilde e_2 e_{p+1},\ldots,\tilde e_1 \tilde e_2 e_{p+q};\tilde e_1 \tilde e_2 e_1,\ldots,\tilde e_1 \tilde e_2 e_p\}, 
\end{align*}
and
\begin{align*}
&Cl_{p,q}\tensor Cl_{0,2}\simeq Cl_{q,p+2},
\end{align*}
i.e.,
\begin{align*}
&\{e_1,\ldots,e_p;e_{p+1},\ldots,e_{p+q}\}\tensor\{;\tilde e_1, \tilde e_2\} \\
&\simeq
\{\tilde e_1 \tilde e_2 e_{p+1},\ldots,\tilde e_1 \tilde e_2 e_{p+q};\tilde e_1, \tilde e_2, \tilde e_1 \tilde e_2 e_1,\ldots,\tilde e_1 \tilde e_2 e_p\}.
\end{align*}
The tensor product with $Cl_{0,2}\simeq\mathbb{R}(2)$
does not change the extension problem.
}
The classifying space for this extension problem is given by
$R_6=Sp(n)/U(n)$, 
with a sufficiently large integer $n$.\cite{kitaev09,morimoto-clifford13}
Since the space of possible representations for $\gamma_z$
is singly connected [$\pi_0(R_6)=0$], one can always find more than one
gamma matrix which can be used as $\gamma_z$ and $\gamma_0$.
This means that a Dirac mass term always exists
so that Weyl/Dirac points can be gapped.
Hence the
instability of Weyl/Dirac points under both time-reversal ($T^2=-1$)
and inversion symmetries known from
Refs.~\onlinecite{Wan-semimetal11,murakami-semimetal07,burkov-balents11}
is understood as the 
existence of a Dirac mass term
which is compatible with the symmetries.

Let us illustrate the instability of a Dirac point with an example.
Suppose that we have a pair of Dirac points,
$\bm{k}=(k_x^0,k_y^0,k_z^0)$ and $(-k_x^0,-k_y^0,-k_z^0)$,
which are related by $T$ and $P$.
The low-energy effective Hamiltonians for the Dirac points are written as
\begin{subequations}
\begin{align}
H_+&=
\sigma_x \tau_y (k_x-k_x^0) + \sigma_y \tau_y (k_y-k_y^0)
 + \sigma_z \tau_y (k_z-k_0),\\
H_-&=
-\sigma_x \tau_y (k_x+k_x^0) - \sigma_y \tau_y (k_y+k_y^0)
 - \sigma_z \tau_y (k_z+k_0),
\end{align}
\end{subequations}
where $\sigma$ and $\tau$ are Pauli matrices representing, e.g., 
spin and orbital degrees of freedom.
With time-reversal and inversion symmetries given by
\begin{align}
T=i \sigma_y \mathcal{K},
\qquad
P=1,
\end{align}
the effective Hamiltonians are transformed as
\begin{subequations}
\begin{align}
TH_\pm(-k_x,-k_y,-k_z)T^{-1}=H_\mp(k_x,k_y,k_z),\\
PH_\pm(-k_x,-k_y,-k_z)P^{-1}=H_\mp(k_x,k_y,k_z),
\end{align}
\end{subequations}
and both $H_+$ and $H_-$ are invariant under the combined transformation,
\begin{align}
PT=i\sigma_y \cal K.
\end{align}
Obviously we can add to $H_\pm$ mass terms
\begin{align}
m_x\tau_x ,~~ m_z\tau_z,
\end{align}
which are invariant under $PT$ and gap out Dirac cones.
Therefore Dirac points are fragile and generally gapped,
in agreement with the general argument based on Clifford algebras.

\subsection{Time-reversal, inversion, and SU(2) spin rotation symmetries:
$\Z_2$ Dirac semimetal}

Let us discuss stability of a Dirac point 
in the presence of time-reversal, inversion,
and SU(2) spin rotation symmetries.
We will demonstrate that the additional SU(2) spin rotation symmetry
completely changes the conclusion of Sec.\ IIIC.
With the SU(2) symmetry, 
we can separate the spin sector and
consider an effective Hamiltonian for spinless fermions.
We thus assume to have symmetry operators satisfying the following relations:
\begin{align}
T^2=+1, \quad
P^2=1, \quad [T,P]=0.
\end{align}
The first equation implies that the system is in class AI.
The combined symmetry operator satisfies
\begin{align}
(PT)^2=+1,
\end{align}
which should be contrasted with Eq.\ (\ref{(PT)^2=-1}).
As we have discussed in Sec.~IIIC, we have a pair of Dirac points,
$\bm{k}=(k_x^0,k_y^0,k_z^0)$ and $(-k_x^0,-k_y^0,-k_z^0)$,
which are related by time-reversal or inversion.
The effective Hamiltonian of a Dirac point is invariant
under $PT$.

Now the Clifford algebra generated from symmetry operators and gamma matrices
reads
\begin{align}
Cl_{0,4} \tensor Cl_{0,2}&=
\{ ; \gamma_x, \gamma_y, \gamma_z, \gamma_0 \} \tensor \{ ; PT, JPT \}.
\end{align}
We can find whether or not a Dirac mass term can exist in the
low-energy effective Hamiltonian of the Dirac point by considering
the following extension problem:
\begin{align}
\{ ; \gamma_x, \gamma_y \} \tensor \{ ; PT, JPT \}
\to \{ ; \gamma_x, \gamma_y, \gamma_z \} \tensor \{ ; PT, JPT \},
\end{align}
\begin{align}
Cl_{0,2} \tensor Cl_{0,2} \to Cl_{0,3} \tensor Cl_{0,2}, 
\end{align}
which is equivalent to $Cl_{0,2}\to Cl_{0,3}$.
The classifying space of this extension problem is
given by $R_2=O(2n)/U(n)$, and its zeroth homotopy group
$\pi_0(R_2)=\mathbb{Z}_2$.
The nontrivial topology of the classifying space indicates that
a Dirac mass term is absent in a minimal Dirac Hamiltonian,
i.e., the massless Dirac Hamiltonian of the least dimensions (4 by 4 matrix)
cannot be gapped out by a Dirac mass term.
However, we can always find a mass term to add to
two copies of minimal models.

For example, let us take 
\begin{align}
T&=\tau_x \cal K,
&
P&=\tau_x , & 
PT&= \cal K,
\end{align}
and write
the Hamiltonian for a Dirac point
\begin{align}
\widetilde H_+=
\sigma_x \tau_z (k_x - k_x^0) + \sigma_z \tau_z (k_y - k_y^0)
 + \tau_x (k_z - k_z^0).
\end{align}
Here Pauli matrices $\sigma_a$, $\tau_b$ are assumed to span the
basis of four orbitals of spinless fermions 
We cannot find any mass term gapping out the Dirac cone
in this 4 by 4 Hamiltonian with preserving $PT$ symmetry.
Thus the gapless Dirac cone is stable when $(PT)^2=+1$.
However, if we double the system by tensoring $\widetilde{H}_+$
with a unit 2 by 2 matrix $\lambda_0$ as
$\widetilde{H}_+ \tensor \lambda_0$,
we can gap out the doubled Dirac cone by
adding mass terms
\begin{align}
\sigma_y \tau_z \lambda_y, \qquad \tau_y \lambda_y,
\end{align}
where $\lambda_x$ and $\lambda_y$ are members of another
set of Pauli matrices $\lambda_\alpha$ ($\alpha=x,y,z$).
Therefore a Dirac point of a minimal (4 by 4) Hamiltonian is stable
while a doubled Dirac point of an 8 by 8 Hamiltonian is unstable,
which indicates that Dirac points are characterized 
by a $\mathbb{Z}_2$ charge.

A lattice regularization of the Dirac Hamiltonian
$\widetilde H_+$ and its time-reversal partner
is given by
\begin{align}
H=& \,
\sigma_x \tau_z \sin k_x + \sigma_z \tau_z \sin k_y \n
&+ \tau_x (\cos k_x+\cos k_y +\cos k_z -M),
\end{align}
with 
symmetry operators
\begin{align}
T&=\tau_x \cal K, & P&= \tau_x.
\end{align}
We have two Dirac points at $(0,0,\pm \pi/2)$ when $M=2$.
These Dirac points are stable.
We note, however, that Dirac points with a nontrivial $\mathbb{Z}_2$ charge 
do not yield helical Fermi arcs,
because the presence of a surface inevitably breaks inversion symmetry.
The bulk-edge correspondence does not hold with inversion symmetry.

\subsection{$\mathbb{Z}_2$ Weyl nodes and $\Z_2$ Dirac nodes in superconductors \label{sec: Z2 nodes}}
In this section we briefly discuss point nodes
in superconductors
that are protected by $\mathbb{Z}_2$ topological charge.
Topological stability of nodes in superconductors
with reflection and inversion symmetries has recently been studied
in Ref.~\onlinecite{kobayashi14}.
Here we focus on two examples that are not discussed
in Ref.~\onlinecite{kobayashi14},
i.e., $\mathbb{Z}_2$ Weyl nodes
and $\mathbb{Z}_2$ Dirac nodes which are 
superconductor analogs of $\mathbb{Z}_2$ Weyl 
and Dirac semimetals.

$\mathbb{Z}_2$ Weyl nodes are stable in the presence of
time-reversal symmetry $T$,
particle-hole symmetry $C$,
and refection symmetry $R_z$ with respect to the $z$ direction.
For a point node at a general $\Vec{k}$ point,
relevant symmetries are $\widetilde T=TR_z$ and $\widetilde C=CR_z$.
We assume symmetry operators satisfy
\begin{align}
\widetilde T^2&=-1, & \widetilde C^2&=+1.
\end{align}
This can be realized in a class DIII superconductor
with ``refection'' symmetry $R_z$,
in which symmetry operators satisfy the relations
\begin{align}
T^2&=-1, & C^2&=+1, & R_z^2&=+1, & [T,R_z]=[C,R_z]&=0.
\end{align}
Again, $R_z$ is a special reflection symmetry that squares to $+1$,
e.g., a combination of reflection and $\pi$ rotation in spin space
as we discussed for $\mathbb{Z}_2$ Weyl semimetals.
Since symmetries impose constraints on Hamiltonian
\begin{subequations}
\label{eq: Z2 Weyl nodes}
\begin{align}
\widetilde T H(-k_x,-k_y,k_z) \widetilde T^{-1}&=H(k_x,k_y,k_z), \\
\widetilde C H(-k_x,-k_y,k_z) \widetilde C^{-1}&=-H(k_x,k_y,k_z),
\end{align}
\end{subequations}
the Hamiltonian $H(k_x,k_y,k_z)$ of fixed $k_z$ can be regarded
as describing a 2D topological superconductor in class DIII, which
is characterized by a $\mathbb{Z}_2$ topological number when
quasiparticle spectra at fixed $k_z$ are fully gapped.
Suppose that the gap closes at some particular
points in the 3D Brillouin zone; these points correspond
to $\mathbb{Z}_2$ topological phase transitions of the fictitious
2D superconductor.
Such gap-closing points are stable and assigned a $\mathbb{Z}_2$
topological charge.
We call them $\mathbb{Z}_2$ Weyl nodes.
Stability of $\mathbb{Z}_2$ Weyl nodes is understood
in terms of Clifford algebra as follows.
We have Clifford algebra for massive Dirac Hamiltonian
with symmetry constraints in Eq.~(\ref{eq: Z2 Weyl nodes}) as
\begin{align}
Cl_{2,5}&=
\{
 J \gamma_x, J \gamma_y, ;
 \widetilde C, J\widetilde C, J\widetilde T \widetilde C, \gamma_z, \gamma_0
\}.
\end{align}
Following the same arguments in the previous subsections
and in the Appendix \ref{app: existence of Dirac mass},
we determine the existence/absence of a Dirac mass term $\gamma_0$
by considering the extension problem
\begin{subequations}
\begin{align}
Cl_{2,3}\to Cl_{2,4}.
\end{align}
Since the classifying space for this is known to be $R_1=O(n)$,
the topological charge of a point node is given by
\begin{align}
\pi_0(R_1)=\mathbb{Z}_2,
\end{align}
\end{subequations}
which reproduces the result of the discussions above.

Next, we discuss $\mathbb{Z}_2$ Dirac nodes that are stable
under the presence of time-reversal symmetry $T$,
particle-hole symmetry $C$, inversion symmetry $P$, 
and SU(2) spin rotation symmetry.
We consider superconductors in class CI,
which is the symmetry class of time-reversal symmetric superconductors
with spin SU(2),\cite{AZ-classes,senthil-fisher98}
and impose additional inversion symmetry.
The three symmetry operators are assumed to satisfy
\begin{subequations}
\begin{align}
T^2&=+1, & C^2&=-1, & P^2&=+1
\end{align}
and
\begin{equation}
[T,C]=[T,P]=[C,P]=0,
\end{equation}
\end{subequations}
where we have assumed even-parity pairing to have $C$ and $P$ commuting
with each other.
Relevant symmetries for a point node are $T'=TP$
and $C'=CP$,
satisfying
\begin{subequations}
\label{eq: Z2 Dirac nodes}
\begin{align}
&T' H(k_x,k_y,k_z) {T'}^{-1}=H(k_x,k_y,k_z), \\
&C' H(k_x,k_y,k_z) {C'}^{-1}=-H(k_x,k_y,k_z), \\
&(T')^2=+1, \qquad  (C')^2=-1.
\end{align}
\end{subequations}

Let us verify that a point node is stable and has
$\mathbb{Z}_2$ topological charge in terms of Clifford algebra.
The Clifford algebra for a massive Dirac Hamiltonian
with symmetry operations $\widetilde T$ and $\widetilde C$ is given by
\begin{align}
Cl_{2,5}&=
\{
 C', JC' ;
 JT' C', \gamma_x, \gamma_y, \gamma_z, \gamma_0
\}.
\end{align}
Then, the existence condition of the Dirac mass term $\gamma_0$
and topological charge of a point node are found from the
following extension problem:
\begin{align}
Cl_{2,3}&\to Cl_{2,4}, & \pi_0(R_1)&=\mathbb{Z}_2.
\end{align}
We thus conclude that point nodes in class CI superconductors
with inversion symmetry are characterized by $\Z_2$ topological charge.

Low-energy effective Hamiltonians for $\mathbb{Z}_2$ Weyl nodes
and $\mathbb{Z}_2$ Dirac nodes
are given by 4 by 4 Bogoliubov-de Gennes (BdG) Hamiltonians.
An example of a BdG Hamiltonian for a pair of
$\mathbb{Z}_2$ Weyl nodes on the $k_z=k_z^0$ plane is given by
\begin{subequations}
\begin{align}
H=k_x \sigma_z \tau_x+ k_y \tau_y + (k_z-k_z^0) \tau_z,
\end{align}
where we have combined the pair of Weyl nodes
by setting $k_x^0=k_y^0=0$.
The relevant symmetry operators [Eq.~(\ref{eq: Z2 Weyl nodes})] are given by
\begin{equation}
\widetilde T=i\sigma_y {\cal K}, \qquad \widetilde C=\tau_x {\cal K},
\end{equation}
\end{subequations}
where $\sigma_\alpha$ and $\tau_\alpha$ are Pauli matrices representing
spin and particle-hole degrees of freedom.
An example of BdG Hamiltonian for a $\mathbb{Z}_2$ Dirac node
at $\Vec k=(k_x^0,k_y^0,k_z^0)$ is given by
\begin{subequations}
\begin{align}
H = (k_x-k_x^0) \sigma_x \tau_x
  + (k_y-k_y^0) \sigma_z \tau_x + (k_z-k_z^0) \tau_z
\end{align}
with the symmetry operators [Eq.~(\ref{eq: Z2 Dirac nodes})]
\begin{align}
T' ={\cal K}, \qquad C'=i\tau_y {\cal K},
\end{align}
\end{subequations}
where $\sigma_\alpha$ and $\tau_\alpha$ are Pauli matrices representing, e.g.,
orbital and particle-hole degrees of freedom.

From the analogy to $\mathbb{Z}_2$ Weyl and Dirac semimetals,
we expect the following features for point nodes with $\Z_2$
topological charge:
$\mathbb{Z}_2$ Weyl nodes appear as a pair of Weyl nodes connected by 
$\widetilde T$, and their projections onto the surface Brillouin zone
are end points of helical Fermi arcs.
A $\mathbb{Z}_2$ Dirac node is not split into a pair of Weyl nodes,
and helical Fermi arcs do not appear in the surface Brillouin zone
because the required inversion symmetry is broken by
the presence of a surface.

\section{Discussions}
In this paper we have proposed Weyl/Dirac semimetals which are characterized
with $\mathbb{Z}_2$ topological charges and
protected by combination of time-reversal symmetry and
additional spatial symmetry:
(a) $\mathbb{Z}_2$ Weyl semimetals protected by time-reversal and
``reflection'' symmetries and
(b) $\mathbb{Z}_2$ Dirac semimetals protected by time-reversal,
inversion, and SU(2) spin rotation symmetries.
The $\mathbb{Z}_2$ Weyl semimetals have helical surface states
forming helical Fermi arcs.
These surface states should give a contribution of $2e^2 k_z^0d/\pi h$
to two-terminal conductance (in analogy to the quantized conductance of
$2e^2/h$ in quantum spin Hall insulators),
where $d$ is the height of the sample in the $z$ direction
and $2k_z^0$ is the separation between two Weyl points
in the $k_z$ direction (perpendicular to the mirror plane)
in the Brillouin zone.%
\footnote{
In addition,
when $S_z$ is conserved in a $\mathbb{Z}_2$ Weyl semimetal protected
by the symmetries in Eq.~(\ref{eq: symmetry T=i sy K and R_z=1}),
helical Fermi arcs contribute $e^2 k^0_z/(\pi h)$ to
spin Hall conductivity, just like
chiral Fermi arcs in conventional Weyl semimetals contribute
$e^2 k^0_z/(\pi h)$ to Hall conductivity\cite{Wan-semimetal11,burkov-balents11}
(which can be described by a $\theta$-term in the effective action of
electromagnetic fields\cite{zyuzin-anomaly12,goswami-tewari13}).}

In the presence of both time reversal symmetry and broken inversion
symmetry, conventional Weyl semimetals are known to appear as an intermediate
phase between a topological insulator phase
and a trivial insulator phase.\cite{murakami-semimetal07}
Similarly, $\mathbb{Z}_2$ Weyl/Dirac semimetals are expected to
appear as an intermediate phase between
a topological insulator phase and a trivial
insulator phase as follows.
When we have time-reversal symmetry $T$ and reflection symmetry $R_z$
[$(TR_z)^2=-1$], we can have 3D topological insulators
with a nontrivial $\mathbb{Z}_2$ topological number
(class AII + $R^+$ in Ref.~\onlinecite{morimoto-clifford13}).
When we have time-reversal, inversion, and spin SU(2) rotation symmetries, 
we can define an integer topological number for 3D gapped phases
(class AI + inversion).\cite{lu-lee14,shiozaki14}
In both cases, at a topological phase transition point where
the topological number changes,
the bulk band gap closes.
Since gap-closing points in these systems are stable thanks to
nontrivial $\mathbb{Z}_2$ charge,
they should remain gapless when a parameter in the Hamiltonian is
changed by a finite amount.
Thus a topological phase transition point
evolves into an intermediate phase of $\mathbb{Z}_2$ Weyl/Dirac semimetals
between a topological insulating phase and a trivial insulating phase.

Finally, we briefly comment on the stability of Weyl/Dirac points against disorder.
What we have shown in Sec.\ III using Clifford algebras is that
Weyl/Dirac points are stable against translation-invariant
perturbations that preserve time-reversal and 
additional spatial symmetries.
On the other hand, disorder is neither translation-invariant 
nor preserves additional spatial symmetry.
Furthermore, disorder can introduce inter-valley scattering
which can gap out Weyl/Dirac points.
However, since potential disorder is irrelevant in the renormalization-group
sense in the three-dimensional bulk,%
\cite{fradkin86,shindou-murakami09}
$\Z_2$ Weyl/Dirac points are expected to be stable against weak disorder.
They should be also stable against weak Coulomb interactions.\cite{goswami-chakravarty11}

$\mathbb{Z}_2$ Weyl semimetals have helical Fermi arcs
connecting projections of Weyl points
onto its surface Brillouin zone.
This is analogous to chiral Fermi arcs in Weyl semimetals.
The chiral surface states of Weyl semimetals are stable
against disorder because of their chiral nature.
On the other hand, in $\Z_2$ Weyl semimetals, random potentials
can induce scattering
among helical surface modes of different $k_z$ and gap them out.
However, if we regard a $\Z_2$ Weyl semimetal as layers of
two-dimensional $\Z_2$ topological insulators labelled by
$k_z$ stacked in momentum space ($-k_z^0<k_z<+k_z^0$),
we can draw analogy to a weak topological insulator which
is layers of two-dimensional $\Z_2$ topological insulators
stacked in real space.
As the surface states of weak topological insulators are
stable against disorder as long as it is spatially uniform
on average,\cite{Ringel12,Mong12,morimoto-weakTI14,Obuse-weakTI13}
we may expect similar stability against disorder
for helical surface modes of $\Z_2$ Weyl semimetals.
Moreover, weak antilocalization effects would drive the surface
to be metallic,
while repulsive Coulomb interactions can alter such metallic surface states into critical states.\cite{ostrovsky10}

\acknowledgments
This work was supported in part by Grants-in-Aid from the Japan Society for
Promotion of Science (Grants No.~24840047 and No.~24540338)
and by the RIKEN iTHES Project.

\appendix*
\section{Existence condition of Dirac mass term \label{app: existence of Dirac mass}}

\begin{table*}[tb]
\begin{center}
\caption{\label{table: AZ classes}
Ten Altland-Zirnbauer symmetry classes and their
topological classification.
Two complex and eight real symmetry classes are characterized
by the presence or the absence of 
time-reversal symmetry ($T$), particle-hole symmetry ($C$) and
chiral symmetry ($\Gamma$).
Their presence is indicated by the sign of squared operator, $T^2$ or $C^2$, 
and by $1$ for $\Gamma$;
their absence is indicated by 0.
For each class, Clifford algebra of $d$ dimensions, the relevant extension problem,
the classifying space $V$, and its zeroth
homotopy group at $d=0$ are listed.
\\
}
\begin{tabular}[t]{cccccccc}
\hline \hline
class & $T$ & $C$ & $\Gamma$ & Clifford algebra & extension & $V$ &
 $\pi_0(V)|_{d=0}$ \\
\hline
A   & 0 & 0    &  0    & $Cl_{d+1}=\{\gamma_0,\gamma_1,\ldots,\gamma_d\}$ & ~$Cl_d \to Cl_{d+1}$ & ~$C_{0+d}$ & $\mathbb{Z}$   \\
AIII  & 0 & 0    &  1  & $Cl_{d+2}=\{\gamma_0,\Gamma,\gamma_1,\ldots,\gamma_d\}$   & ~$Cl_{d+1} \to Cl_{d+2}$ & ~$C_{1+d}$ & 0 \\
\hline
AI   & $+1$~ & 0    &  0  & $Cl_{1,d+2}=\{J\gamma_0; T, TJ, \gamma_1,\ldots,\gamma_d\}$ & ~$Cl_{0,d+2} \to Cl_{1,d+2}$ & ~$R_{0-d}$ & $\mathbb{Z}$   \\
BDI  & $+1$~ & $+1$~ &  1  & $Cl_{d+1,3}=\{J\gamma_1,\ldots,J\gamma_d,TCJ; C,CJ,\gamma_0\}$   & ~$Cl_{d+1,2} \to Cl_{d+1,3}$ & ~$R_{1-d}$ & $\mathbb{Z}_2$ \\
D    & 0    & $+1$~ &  0  & $Cl_{d,3}=\{J\gamma_1,\ldots,J\gamma_d; C,CJ,\gamma_0\}$   & ~$Cl_{d,2} \to Cl_{d,3}$   & ~$R_{2-d}$ & $\mathbb{Z}_2$ \\
DIII & $-1$~ & $+1$~ &  1  & $Cl_{d,4}=\{J\gamma_1,\ldots,J\gamma_d; C,CJ,TCJ,\gamma_0\}$   & ~$Cl_{d,3} \to Cl_{d,4}$   & ~$R_{3-d}$ & 0              \\
AII  & $-1$~ & 0    &  0  & $Cl_{3,d}=\{J\gamma_0, T, TJ; \gamma_1,\ldots, \gamma_d \}$   & ~$Cl_{2,d} \to Cl_{3,d}$   & ~$R_{4-d}$ & $\mathbb{Z}$   \\
CII  & $-1$~ & $-1$~ &  1  & $Cl_{d+3,1}=\{J\gamma_1,\ldots,J\gamma_d,C,CJ,TCJ; \gamma_0\}$   & ~$Cl_{d+3,0} \to Cl_{d+3,1}$ & ~$R_{5-d}$ & 0              \\
C    & 0    & $-1$~ &  0  & $Cl_{d+2,1}=\{J\gamma_1,\ldots,J\gamma_d, C, CJ; \gamma_0\}$   & ~$Cl_{d+2,0} \to Cl_{d+2,1}$ & ~$R_{6-d}$ & 0              \\
CI   & $+1$~ & $-1$~ &  1  & ~$Cl_{d+2,2}=\{J\gamma_1,\ldots,J\gamma_d,C,CJ; TCJ,\gamma_0\}$~   & ~$Cl_{d+2,1} \to Cl_{d+2,2}$ & ~$R_{7-d}$ & 0              \\
\hline \hline
\end{tabular}
\end{center}
\end{table*}

Considering the extension problem of Clifford algebras,
we can tell whether we can add a Dirac mass term 
to a given massless Dirac Hamiltonian under symmetry constraints.
In this Appendix we discuss existence conditions of Dirac mass 
for ten Altland-Zirnbauer symmetry classes.
This is based on the following idea:
\begin{itemize}
\item 
In the classification scheme with Clifford algebras, the
existence condition of a particular generator $e_i$ (Dirac mass term)
is equivalent to classification of 
another generator of the same type in Clifford algebra
in which $e_i$ is removed.
\end{itemize}

First, let us briefly review classification of
massive Dirac Hamiltonians using Clifford algebras
(for details, see Ref.~\onlinecite{morimoto-clifford13}).
Table \ref{table: AZ classes} summarizes the result of classification
for a massive Dirac Hamiltonian in $d$ dimensions, 
\begin{align}
H=\sum_{i=1}^d k_i \gamma_i + m \gamma_0,
\label{eq: H appendix}
\end{align}
where $\gamma_j$ $(j=0,1,\ldots,d)$ are gamma matrices.
$H$ belongs to one of the Altland-Zirnbauer symmetry class
which is specified by the presence or absence of three generic symmetries:
time-reversal symmetry $T$, particle-hole symmetry $C$,
and chiral symmetry $\Gamma$.
A set of gamma matrices $(\gamma_j)$ and symmetry operators
($\Gamma$ in class AIII; $T$ and/or $C$ and imaginary unit $J$ in real classes)
form Clifford algebra as shown in Table \ref{table: AZ classes}.
By examining the extension problem with respect to the Dirac mass term,
we can obtain classifying space $V$ which is a space of all possible
Dirac mass terms under symmetry constraints.
Then the topological classification is found from
its zeroth homotopy group $\pi_0(V)$
[the last column in Table \ref{table: AZ classes} lists
$\pi_0(V)$ for $0$-dimensional systems].

The type of topological indices $(\mathbb{Z},\mathbb{Z}_2,0)$ 
characterizing massive Dirac Hamiltonians determines
whether we have a unique Dirac mass $\gamma_0$
or we have multiple Dirac masses that anticommute with each other,
as we explain below.
That is, topology of classifying space can be used
to understand uniqueness/multipleness of the Dirac mass term.
When the Dirac Hamiltonian $H$ has
only a single Dirac mass term $m\gamma_0$ which is allowed by
assumed symmetry constraints of the symmetry class,
the ground state of $H$ for $m>0$ and that for $m<0$ are
topologically distinct, because they cannot be connected without
closing the bulk gap $m$.
This corresponds to the case when the zeroth homotopy group of
the classifying space $V$ is nontrivial, i.e., $\mathbb{Z}$ or $\mathbb{Z}_2$.
The difference between $\mathbb{Z}$ and $\mathbb{Z}_2$ manifests itself
if we double the system, $H\otimes \sigma_0$, 
where $\sigma_0$ is a $2\times2$ identity matrix.
For the $\mathbb{Z}_2$ case,
we can find an extra mass term $m'\gamma_0'$ that anticommutes with
$H\otimes\sigma_0$ (note that $m\gamma_0$ is included in $H$).
Then the ground states of $H\otimes\sigma_0$ with different signs of
the mass $m$ are no longer topologically distinct,
since we can adiabatically deform the Dirac mass term as
$m\gamma_0\otimes\sigma_0\cos\theta+m'\gamma_0'\sin\theta$
($0\le\theta\le\pi$).
On the other hand, when the zeroth homotopy of the classifying space $V$
is $\mathbb{Z}$,
we cannot find any extra mass term that anticommutes with $H\otimes\sigma_0$,
because two copies of topologically non-trivial systems add up
and the states with different signs of the mass $m$ are still distinct.
When $H$ has more than one Dirac mass terms,
the gapped ground states of $H$ can be adiabatically connected
without closing the energy gap.
For example, if $H$ has two Dirac mass terms,
$m\gamma_0=m_1\gamma_{0,1}+m_2\gamma_{0,2}$ with
$\{\gamma_{0,1},\gamma_{0,2}\}=0$,
then the ground states of $H$ with $m\gamma_0=+m\gamma_{0,1}$
and $m\gamma_0=-m\gamma_{0,1}$ are not topologically distinct,
since we can connect them by the homotopy 
\begin{align}
\gamma_0(\theta)=\cos \theta \gamma_{0,1} + \sin \theta \gamma_{0,2},
~~(0\le \theta \le \pi). 
\label{path}
\end{align}
In this case the classification of the symmetry class
is trivial, $\pi_0(V)=0$.

Now let us turn to the existence condition of the Dirac mass term $\gamma_0$
for given kinetic gamma matrices and symmetry constraints.
Suppose that the extension problem with respect to the mass term $\gamma_0$
of the Dirac Hamiltonian [Eq.~(\ref{eq: H appendix})]
has the form
\begin{align}
&Cl_{p,q}=\{e_1,\ldots,e_p; e_{p+1},\ldots,e_{p+q}\} \n
&\to Cl_{p,q+1}=\{e_1,\ldots,e_p; e_{p+1},\ldots,e_{p+q},\gamma_0\},
\label{eq: classification of gamma}
\end{align}
the relevant classifying space is $R_{q-p}$.
(This example corresponds to symmetry classes with particle-hole symmetry;
see Table~\ref{table: AZ classes}.)
The existence of $\gamma_0$ is then determined by the extension problem
with one less generator,
\begin{align}
&Cl_{p,q-1}=\{e_1,\ldots,e_p; e_{p+1},\ldots,e_{p+q-1}\} \n
&\to Cl_{p,q}=\{e_1,\ldots,e_p ; e_{p+1}, \ldots,e_{p+q-1},e_{p+q}\}.
\label{eq: existence condition}
\end{align}
If we denote the classifying space for this extension problem by
$\widetilde V$, then $\widetilde V=R_{q-p-1}$.
Notice the change in the index of the classifying space by $-1$.
As we have seen,
topology of the classifying space $R_{q-p-1}$ for the extension
problem of the generator $e_{p+q}$ tells us whether
$e_{p+q}$ is unique or not, i.e., 
whether there exists an extra operator $\tilde e_{p+q}$ that is the same type
as $e_{p+q}$ and anticommutes with $e_{p+q}$.
Since the extra operator $\tilde e_{p+q}$ can be adopted as 
a Dirac mass term $\gamma_0$,
uniqueness/multipleness of the operator $e_{p+q}$ 
corresponds exactly to absence/presence of the Dirac mass term $\gamma_0$ as follows.

If $\pi_0(\widetilde V)=\mathbb{Z}$,
we cannot find any extra operator that anticommutes with
the generators $e_1,\ldots,e_{p+q}$ and squares to $+1$;
hence $\gamma_0$ does not exist.
If $\pi_0(\widetilde V)=\mathbb{Z}_2$, 
the existence of $\gamma_0$ depends on
the size of the Dirac Hamiltonian that we consider.
When a minimal Dirac Hamiltonian under given symmetry constraints
has the matrix form of dimension $n$,
the dimension of general Dirac Hamiltonians with the same symmetries is
given by $kn$, where $k$ is an integer.
A mass term $\gamma_0$ can be present in Dirac Hamiltonians of $k$ even, 
while it cannot be present in Dirac Hamiltonians of $k$ odd.
Finally, if $\pi_0(\widetilde V)=0$,
we can always find an extra generator, i.e., $\gamma_0$ exists.

We can repeat the same discussion for class AI and AII.
For these classes the extension problem with respect to $J\gamma_0$ is
of the form $Cl_{p,q}\to Cl_{p+1,q}$,
whose classifying space is $V=R_{p-q+2}$
(see Table~\ref{table: AZ classes}).
The extension problem with one less generator similar to
Eq.\ (\ref{eq: existence condition}) is
$Cl_{p-1,q}\to Cl_{p,q}$, for which the classifying space is
$\widetilde V=R_{p-q+1}$
(note the change in the index by $-1$).
The existence of $\gamma_0$ is judged from $\pi_0(\widetilde V)$.

Finally, the existence condition of $\gamma_0$ for complex classes A and AIII
is obtained by replacing real Clifford algebras in
Eq.~(\ref{eq: existence condition}) with complex algebras, i.e.,
$Cl_{q-1}\to Cl_q$, where $q=d$ for class A and $q=d+1$ for class AIII.

In summary, 
when the classifying space for Eq.~(\ref{eq: classification of gamma})
is $V=R_q$ ($C_q$),
the classifying space for Eq.~(\ref{eq: existence condition})
is given by $\widetilde V=R_{q-1}$ ($C_{q-1}$).
Depending on the topology of $\widetilde V$,
we have the following three cases regarding the existence
of a Dirac mass term $\gamma_0$ in Dirac Hamiltonian of $kn$ dimensions,
where $n$ is the minimal size of Dirac Hamiltonians in a given set of
symmetry constraints:
\begin{itemize}
\item $\pi_0(\widetilde V)=\mathbb{Z}$: 
No Dirac mass term $\gamma_0$ exists for any integer $k$.

\item $\pi_0(\widetilde V)=\mathbb{Z}_2$:
No Dirac mass term $\gamma_0$ exists for odd $k$,
while $\gamma_0$ can exist for even $k$.

\item $\pi_0(\widetilde V)=0$:
At least one Dirac mass term $\gamma_0$ can be found for any $k$.
\end{itemize}

We note that, for each symmetry class,
the existence condition of a Dirac mass term in $d$-dimensional Dirac
Hamiltonian is directly related to the classification of topological
insulators/superconductors in the same symmetry class in $d+1$ dimensions.
This can be seen by noticing that the change in the index $q$ of
the classifying space $R_q$ by $-1$ ($C_q$ by $-1=+1$ mod $2$) is equivalent to
increasing the space dimension $d$ by $+1$ in Table~\ref{table: AZ classes}.
For example, if a $d$-dimensional system is a boundary of
a topological insulator/superconductor in $d+1$ dimensions, then
the nontrivial boundary states cannot be gapped.
Naturally, this indicates that there is no Dirac mass term for the
gapless Dirac fermions on the $d$-dimensional surface of a
$(d+1)$-dimensional topological insulator/superconductor.

We also note that the existence condition of Dirac mass terms
discussed in this Appendix
gives a topological charge of gap-closing points 
located at time-reversal invariant momenta in the
ten Altland-Zirnbauer symmetry classes.
However, when gap-closing points are not located on
time-reversal invariant momenta,
their topological charge is related to the existence condition
of a complex class (A or AIII),
because time-reversal and particle-hole symmetries are not symmetries
of a Dirac Hamiltonian for a single gap-closing point.

%

\end{document}